\documentclass[aps,preprint,showpacs,preprintnumbers,amsmath,amssymb]{revtex4}

\usepackage{amsmath,mathrsfs,amsbsy,color,graphicx,bm,amsthm,amsfonts}
\usepackage{units}
\usepackage{bbm}
\usepackage{times}
\usepackage{dcolumn}
\usepackage{mathrsfs}
\usepackage{amsmath,amssymb,epsfig}
\usepackage{amsmath}
\newcommand{\udots}{\mathinner{\mskip1mu\raise1pt\vbox{\kern7pt\hbox{.}}
\mskip2mu\raise4pt\hbox{.}\mskip2mu\raise7pt\hbox{.}\mskip1mu}}

\begin{document}

\title{Sharing of Gaussian tripartite steering in de-Sitter space}
\author{Xin-Ze Song$^1$, Xiao-Li Huang$^1$\footnote{huangxiaoli1982@foxmail.com}, Shu-Min Wu$^1$\footnote{smwu@lnnu.edu.cn}}
\affiliation{Department of Physics, Liaoning Normal University, Dalian 116029, China}

\begin{abstract}
We investigate the redistribution and directional properties of Gaussian tripartite quantum steering in de-Sitter space within the continuous-variable framework. By expressing the Bunch-Davies vacuum in terms of the open-chart modes through a Bogoliubov transformation, we derive the covariance matrix of the resulting multipartite Gaussian state and analyze both one-to-two and two-to-one steering configurations. We find that
de-Sitter curvature generally suppresses the steering initially shared among the accessible modes, but its influence depends strongly on the steering direction. In particular, while some steering configurations undergo sudden death at finite curvature, two-to-one steering toward a nongravitational mode remains finite
throughout the parameter range considered, demonstrating a robustness of tripartite steering absent in its bipartite counterpart. Remarkably, the maximum steering asymmetry occurs at the critical point at which steering in one direction undergoes sudden death, marking the transition from two-way to one-way steering. We further show that the de-Sitter gravitational field can generate tripartite steering involving modes separated by the cosmological horizon. Depending on the steering partition, this curvature-induced steering exhibits persistent one-way or asymmetric two-way behavior, as well as a sudden-birth transition from one-way to two-way steering. These results demonstrate that de-Sitter curvature not only degrades preexisting quantum correlations but also redistributes and generates directional multipartite steering across causally disconnected regions.
\end{abstract}

\vspace*{0.5cm}
\maketitle

\section{Introduction}
Einstein–Podolsky–Rosen (EPR) steering characterizes the ability to nonclassically influence a remote quantum state through local measurements and represents an intermediate form of quantum correlation between entanglement and Bell nonlocality \cite{L1,L2,L3,L4}. As quantum information processing evolves from simple bipartite systems to multipartite and network-based scenarios, bipartite steering becomes insufficient to meet the demands of complex tasks. Consequently, multipartite steering has attracted increasing attention due to its potential applications in multipartite state distribution, resource sharing, and distributed quantum processing \cite{L5,L6,L7}. In multipartite systems, steering can exhibit various structures, such as one-to-many and many-to-one configurations \cite{L8,L9}, providing more versatile mechanisms for information manipulation among multiple nodes. From the perspective of quantum information theory, EPR steering is inherently directional and can be classified into one-way and two-way steering. The former refers to situations where steering exists only in a single direction while the reverse direction is absent \cite{L10}, whereas the latter corresponds to mutual steerability between parties. This directional asymmetry not only enriches the hierarchy of quantum correlations but has also been experimentally demonstrated \cite{L11,L12}, providing practical support for one-sided device-independent quantum information protocols \cite{L13}.

Relativistic quantum information provides a framework for exploring how relativistic motion and gravitational fields modify quantum states and quantum-information resources \cite{SDF1,SDF2,SDF3,SDF4,SDF5,SDF6,SDF7,SDF8,SDF9,SDF10,SDF11,SDF12,SDF13,SDF14,SDF15,SDF16,SDF17,SDF18,SDF19,SDF20,SDF21,SDF22,SDF23,SDF24,SDF25,SDF26,SDF27,SDF28,SDF29,SDF30,SDF31,SDF32,SDF33,SDF34,SDF35,SDF36,SDF37,SDF38,SDF39,SDF40,SDF41,SDF42,SDF43,SDF44,SDF45,SDF46,SDF47,SDF48,SDF49,SDF50,SDF51,SDF52,
SDF53,SDF54,SDF55,SDF56,SDF57,AGL1,AGL2}. These effects become particularly significant in curved space with causal horizons, where the restriction of observers to finite causal regions renders part of the field degrees of freedom inaccessible \cite{SDF1,SDF2,SDF3,SDF4,SDF5,SDF6,SDF7,SDF8,SDF9,SDF10,SDF11,SDF12,SDF13,SDF14,SDF15}. De-Sitter space, characterized by exponential expansion and a cosmological event horizon, provides a natural setting for investigating such horizon-induced modifications of quantum correlations. The horizon divides the spacetime into two causally disconnected regions, conventionally denoted by $R$ and $L$, while the Bunch-Davies vacuum exhibits intrinsic quantum correlations between field modes across these regions \cite{ACP1}. Through a Bogoliubov transformation, the global field modes can be related to modes localized in the two open charts, allowing the Bunch-Davies vacuum to be represented as a two-mode squeezed state \cite{ACP2,ACP3,ACP4,ACP5}. For an observer confined to one causal region, the modes beyond the horizon are inaccessible and must be traced out, converting an initially pure state into a mixed state and consequently modifying the observable quantum resources \cite{ACP6,ACP7,ACP8}. Continuous-variable Gaussian states provide a particularly suitable framework for characterizing these effects, since their quantum correlations can be systematically analyzed in phase space through covariance matrices. Previous studies of quantum correlations in de-Sitter space have mainly focused on bipartite systems, where quantum steering, entanglement, and quantum discord can be strongly degraded by spacetime curvature and may even vanish within finite parameter regimes \cite{ACP9,ACP10,ACP11,ACP12}. Such behavior demonstrates the fragility of two-mode quantum correlations in the presence of a cosmological horizon. Multipartite systems, however, possess a substantially richer correlation structure: quantum resources can be shared and redistributed among different subsystems, and distinct one-to-two and two-to-one steering configurations can respond differently to the same gravitational environment. This raises several fundamental questions. Can multipartite steering survive in regimes where its bipartite counterpart disappears? How does de-Sitter curvature affect the asymmetry between opposite steering directions? More importantly, can the correlations induced across the cosmological horizon generate multipartite steering involving causally disconnected modes? Addressing these questions is essential for understanding how spacetime curvature reshapes the distribution and directionality of multipartite quantum resources.

In this work, we investigate the sharing and directional properties of Gaussian tripartite quantum steering in de-Sitter space within the continuous-variable framework. We consider an initially correlated three-mode Gaussian state shared by Alice ($A$), Bob ($B$), and Charlie ($C$). Owing to the presence of the cosmological horizon, Charlie's mode is related through the Bogoliubov transformation to an additional mode $\bar{C}$ located in the causally disconnected region $L$, leading to an effective four-mode description. By expressing the Bunch-Davies vacuum in terms of the open-chart modes, we derive the corresponding covariance matrices and characterize the evolution of Gaussian tripartite steering in phase space. We focus on two complementary aspects of the problem. First, we examine how the de-Sitter curvature and the field mass affect different steering partitions among the physically accessible modes, including $A\rightarrow BC$, $BC\rightarrow A$, $AB\rightarrow C$, and $C\rightarrow AB$. These configurations allow us to compare one-to-two and two-to-one steering and to explore how spacetime curvature modifies their directional asymmetry and robustness. Second, we extend the analysis to steering configurations involving the causally inaccessible mode $\bar{C}$, with particular attention to the redistribution of quantum steering across the cosmological horizon. Through these complementary scenarios, we explore how the causal structure and gravitational effects of de-Sitter space reshape the sharing, directionality, and redistribution of multipartite quantum steering. As will be demonstrated, Gaussian tripartite steering exhibits a pronounced dependence on the steering direction and reveals rich steering characteristics arising from the interplay between multipartite correlations and de-Sitter gravitational effects.

The remainder of this paper is organized as follows. In Sec.~II, we review 
the quantization of a free scalar field in de-Sitter space and the 
Bogoliubov transformation relating the Bunch-Davies vacuum to the open-chart 
vacua. In Sec.~III, we introduce the Gaussian steering measure for 
continuous-variable systems. In Sec.~IV, we investigate the sharing and 
directional asymmetry of Gaussian tripartite steering in de-Sitter space, 
considering steering configurations within the physically accessible region 
as well as those involving the causally disconnected region. Finally, 
Sec.~V summarizes our main results and conclusions.

\section{Quantization of a scalar field in de-Sitter space}

\begin{figure}[h]
\centering
\includegraphics[width=0.7\textwidth]{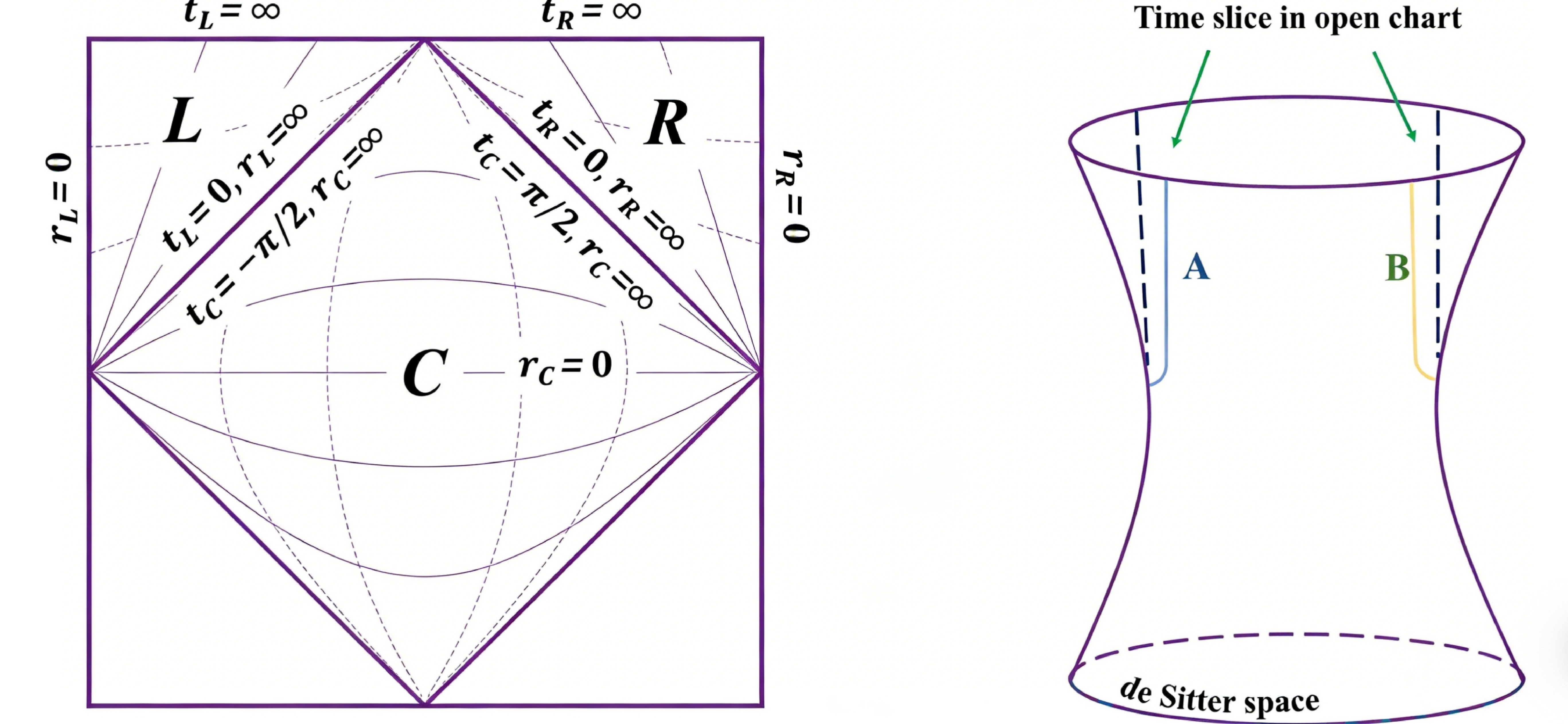}
\caption
{The Penrose diagram of the de-Sitter space, in which $L$ and $R$
are the two causally disconnected regions described by the open charts.}
\label{O1}
\end{figure}
Consider a free massive scalar field \(\phi\) with effective mass \(m\) propagating in de-Sitter space described by the metric \(g_{\mu\nu}\) \cite{ACP1}. The corresponding action is given by
\begin{align}\label{A1}
S=\int d^{4}x\sqrt{-g}\left[-\frac{1}{2}g^{\mu\nu}\partial_{\mu}\phi\partial_{\nu}\phi-\frac{m^{2}}{2}\phi^{2}\right].
\end{align}
In the open-chart coordinates of de-Sitter space, the metrics in the \(R\) and \(L\) regions shown in Fig.~{\ref{O1}} can be obtained through analytic continuation of the Euclidean metric:
\begin{align}\label{A2}
ds^{2}_{R}=H^{-2}\left[-dt^{2}_{R}+\sinh^{2}t_{R}\left(dr^{2}_{R}+\sinh^{2}r_{R}d\Omega^{2}\right)\right],\nonumber\\
ds^{2}_{L}=H^{-2}\left[-dt^{2}_{L}+\sinh^{2}t_{L}\left(dr^{2}_{L}+\sinh^{2}r_{L}d\Omega^{2}\right)\right].
\end{align}
Here, \(H^{-1}\) denotes the Hubble radius, and \(d\Omega^{2}\) is the metric on the unit two-sphere. The solutions of the Klein-Gordon equation can be separated into temporal and spatial parts and written as
\begin{align}\label{A3}
u_{\sigma p \ell m}(t,r,\Omega)
\sim\frac{H}{\sinh t}\chi_{p,\sigma}(t)Y_{p\ell m}(r,\Omega),
\end{align}
where \(Y_{p\ell m}(r,\Omega)\) are the harmonic functions defined on the three-dimensional hyperbolic spatial hypersurfaces and satisfy
\begin{align}\label{A4}
-L^{2}Y_{p\ell m}=(1+p^{2})Y_{p\ell m}.
\end{align}
The parameter \(p\) characterizes the hyperbolic momentum of the field modes, while \(\ell\) and \(m\) denote the angular quantum numbers.

The temporal functions \(\chi_{p,\sigma}\) are the positive-frequency mode functions associated with the Bunch-Davies vacuum \cite{ACP1}. Their expressions in the \(R\) and \(L\) regions are given by
\begin{align}\label{A5}
\chi_{p,\sigma}(t_{R(L)})=\left\{
\begin{aligned}
&
\frac{e^{\pi p}- i\sigma e^{-i\pi \nu}}{\Gamma\!\left(\nu + i p + \tfrac12\right)}\, P^{i p}_{\nu - \tfrac12}(\cosh t_R)-\frac{e^{-\pi p}- i\sigma e^{-i\pi \nu}}{\Gamma\!\left(\nu - i p + \tfrac12\right)}\, P^{-i p}_{\nu - \tfrac12}(\cosh t_R)
\\[1.2em]
&
\frac{\sigma e^{\pi p}- i e^{-i\pi\nu}}{\Gamma\!\left(\nu + i p + \tfrac12\right)}\, P^{i p}_{\nu - \tfrac12}(\cosh t_L)-\frac{\sigma e^{-\pi p}- i e^{-i\pi\nu}}{\Gamma\!\left(\nu - i p + \tfrac12\right)}\, P^{-i p}_{\nu - \tfrac12}(\cosh t_L),
\end{aligned}
\right.
\end{align}
where \(P^{\pm ip}_{\nu-\frac12}\) are the associated Legendre functions. The index \(\sigma=\pm1\) distinguishes the two independent solutions. Returning to the temporal mode functions \(\chi_{p,\sigma}\), the Klein-Gordon normalization condition determines the normalization factor as
\begin{align}\label{A6}
N_{p}=\frac{4\sinh(\pi p)\sqrt{\cosh(\pi p)-\sigma\sin(\pi \nu)}}{\sqrt{\pi}\left|\Gamma\left(\nu+ip+\tfrac{1}{2}\right)\right|}.
\end{align}
In the above solutions, the mass parameter \(\nu\) is defined by
\begin{align}\label{A7}
\nu=\sqrt{\frac{9}{4}-\frac{m^{2}}{H^{2}}}.
\end{align}
The parameter \(p\) also characterizes the sensitivity of the field modes to the curvature of the open spatial hypersurfaces. In our setting, the curvature effects become stronger as \(p\) approaches zero, whereas they become relatively weaker as \(p\) approaches unity. Two particularly important values of the mass parameter are \(\nu=\tfrac12\) and \(\nu=\tfrac32\). In the usual effective-mass description, \(\nu=\tfrac12\) corresponds to the conformally coupled massless scalar field, for which the effective mass satisfies \(m^{2}=2H^{2}\), whereas \(\nu=\tfrac32\) corresponds to the minimally coupled massless limit.

The scalar field operator can then be expanded in terms of these modes as
\begin{align}\label{A8}
\widehat{\phi}(t,r,\Omega)
=
\frac{H}{\sinh t}
\int dp
\sum_{\ell,m}
\phi_{p\ell m}(t)
Y_{p\ell m}(r,\Omega),
\end{align}
where the time-dependent operator-valued mode coefficient is defined by
\begin{align}\label{A9}
\widehat{\phi}_{p\ell m}(t)
\equiv
\sum_{\sigma}
\left[
a_{\sigma p\ell m}\chi_{p,\sigma}(t)
+
a_{\sigma p\ell,-m}^{\dagger}
\chi_{p,\sigma}^{*}(t)
\right].
\end{align}
Here, $a_{\sigma p\ell m}$ and $a_{\sigma p\ell m}^{\dagger}$ are the annihilation and creation operators associated with the Bunch-Davies modes $\chi_{p,\sigma}(t)$, respectively. Their nonvanishing canonical commutation relation is
\begin{align}\label{A10}
\left[
a_{\sigma p\ell m},
a_{\sigma' p'\ell' m'}^{\dagger}
\right]
=
\delta(p-p')
\delta_{\sigma\sigma'}
\delta_{\ell\ell'}
\delta_{mm'},
\end{align}
while all commutators between two annihilation operators or two creation operators vanish. The Bunch-Davies vacuum is therefore defined as the state annihilated by all $a_{\sigma p\ell m}$:
\begin{align}\label{A11}
a_{\sigma p\ell m}|0\rangle_{BD}=0.
\end{align}
For simplicity, we omit the indices \(p\), \(\ell\), and \(m\) in \(\phi_{p\ell m}\), \(a_{\sigma p\ell m}\), and \(a^{\dagger}_{\sigma p\ell m}\) hereafter. We also introduce the shorthand notation
$$
\chi_{p,\sigma}(t)\rightarrow\chi^{\sigma},\qquad
P^{ip}_{\nu-\frac{1}{2}}(\cosh t_{R,L})\rightarrow P^{R,L},
\qquad
P^{-ip}_{\nu-\frac{1}{2}}(\cosh t_{R,L})\rightarrow P^{R,L\,\ast}.
$$
Next, we introduce positive-frequency mode functions associated with the \(R\) and \(L\) vacua, which have support only in the corresponding regions. They are defined as
\begin{align}\label{A12}
\varphi^{q}=\left\{
\begin{aligned}
&\tilde{N}_{p}^{-1} P^{q}, && \text{in region } q\\
&0, && \text{in the opposite region},
\end{aligned}
\right.
\end{align}
where \(q=(R,L)\) and
\begin{align}\label{A13}
\tilde{N}_{p}=\frac{\sqrt{2p}}{\left|\Gamma(1+ip)\right|}.
\end{align}
Since the field operator is independent of the choice of mode basis, it can equivalently be expanded in terms of the Bunch-Davies modes or the modes localized in the \(R\) and \(L\) regions:
\begin{align}\label{A14}
\phi(t)=a_{\sigma}\chi^{\sigma}+a^{\dagger}_{\sigma}\chi^{\sigma\ast}
=b_{q}\varphi^{q}+b^{\dagger}_{q}\varphi^{q\ast},
\end{align}
where \(b_q\) and \(b_q^{\dagger}\) are the annihilation and creation operators associated with the local modes and satisfy
\(b_q|0\rangle_q=0\). The two sets of operators are therefore related through a Bogoliubov transformation.

As a consequence, the Bunch-Davies vacuum is not a direct product of the \(R\)- and \(L\)-region vacua  \cite{ACP2,ACP3,ACP4}. Instead, it can be expressed as a squeezed state in the tensor-product Hilbert space \(\mathcal{H}_{R}\otimes\mathcal{H}_{L}\):
\begin{align}\label{A15}
|0\rangle_{BD}\propto\exp\left(\frac{1}{2}\sum_{i,j=R,L} m_{ij} b^{\dagger}_{i}b^{\dagger}_{j}\right)|0\rangle_{R}|0\rangle_{L}.
\end{align}
The symmetric matrix \(m_{ij}\) is determined by imposing the condition \(a_{\sigma}|0\rangle_{BD}=0\), yielding
\begin{align}\label{A16}
m_{ij}=e^{i\theta}\frac{\sqrt{2}e^{-p\pi}}{\sqrt{\cosh(2\pi p)+\cos(2\pi \nu)}}
\begin{pmatrix}
\cos(\pi\nu) & i\sinh(\pi p)\\
i\sinh(\pi p) & \cos(\pi\nu)
\end{pmatrix}.
\end{align}
Here, \(e^{i\theta}\) represents an unphysical phase factor for \(\nu^{2}>0\). For general values of \(\nu\), the diagonal elements \(m_{RR}=m_{LL}\) are nonzero, indicating the presence of local squeezing within the \(R\) and \(L\) sectors in addition to the correlations between them. This makes it inconvenient to obtain the reduced density matrix directly in the original \(b_R\) and \(b_L\) number bases. For the special cases \(\nu=\tfrac12\) and \(\nu=\tfrac32\), however, \(\cos(\pi\nu)=0\), so that the diagonal elements of \(m_{ij}\) vanish and the squeezed-state structure becomes considerably simpler.

To obtain a form that is more convenient for tracing over either region, we perform an additional Bogoliubov transformation separately within the \(R\) and \(L\) sectors. This transformation does not mix the Hilbert spaces \(\mathcal{H}_{R}\) and \(\mathcal{H}_{L}\). We introduce the new operators
\begin{align}\label{A17}
c_{R}=u b_{R}+v b^{\dagger}_{R},
\qquad
c_{L}=u^{\ast} b_{L}+v^{\ast} b^{\dagger}_{L}.
\end{align}
The condition $|u|^{2}-|v|^{2}=1$ ensures that the transformed operators obey the canonical commutation relations
\([c_i,c_j^{\dagger}]=\delta_{ij}\). In terms of these operators, the Bunch-Davies vacuum takes the standard two-mode squeezed form
\begin{align}\label{A18}
|0\rangle_{BD}=N^{-1}_{\gamma_{p}}\exp\left(\gamma_{p}c^{\dagger}_{R}c^{\dagger}_{L}\right)|0\rangle_{R'}|0\rangle_{L'}.
\end{align}
Here, \(|0\rangle_{R'}\) and \(|0\rangle_{L'}\) denote the vacua associated with the transformed operators \(c_R\) and \(c_L\), respectively. For simplicity, we henceforth omit the primes and denote the transformed modes \(R'\) and \(L'\) simply by \(R\) and \(L\). Accordingly, \(|0\rangle_R\) and \(|0\rangle_L\) in the following refer to the vacua of the transformed operators \(c_R\) and \(c_L\). The normalization factor is given by
\begin{align}\label{A19}
N_{\gamma_{p}}^{2}
=\left|\exp\left(\gamma_{p} c^{\dagger}_{R} c^{\dagger}_{L}\right)|0\rangle_{R'}|0\rangle_{L'}\right|^{2}
=\frac{1}{1-|\gamma_{p}|^{2}},
\end{align}
where the condition \(|\gamma_{p}|<1\) must be satisfied. Using the definitions of \(c_R\) and \(c_L\), together with the two-mode squeezed form of the Bunch-Davies vacuum, one obtains $c_{R}|0\rangle_{BD}=\gamma_{p} c^{\dagger}_{L}|0\rangle_{BD},
c_{L}|0\rangle_{BD}=\gamma_{p} c^{\dagger}_{R}|0\rangle_{BD}.$
Defining
\(\omega\equiv m_{RR}=m_{LL}\) and
\(\zeta\equiv m_{RL}=m_{LR}\), these consistency relations lead to
\begin{align}\label{A20}
\gamma_{p}=\frac{1}{2\zeta}\left[-\omega^{2}+\zeta^{2}+1-\sqrt{(\omega^{2}-\zeta^{2}-1)^{2}-4\zeta^{2}}\right].
\end{align}
Substituting the matrix elements of \(m_{ij}\) into this expression, we obtain
\begin{align}\label{A21}
\gamma_{p}=i\frac{\sqrt{2}}{\sqrt{\cosh(2\pi p)+\cos(2\pi \nu)}+\sqrt{\cosh(2\pi p)+\cos(2\pi \nu)+2}}.
\end{align}
Thus, the squeezing parameter \(\gamma_p\), and consequently the correlations between the \(R\) and \(L\) regions, depend on both the hyperbolic momentum \(p\) and the mass parameter \(\nu\). For the conformally coupled scalar field with \(\nu=\tfrac12\) and the minimally coupled massless scalar field with \(\nu=\tfrac32\), the expression reduces to
$|\gamma_{p}| = e^{-\pi p}$.

\section{Quantification of Gaussian quantum steering}

For a two-mode Gaussian state $\rho_{AB}$ shared by Alice and Bob, we introduce the quadrature-operator vector $\hat{R}=(\hat{x}^{A},\hat{p}^{A},\hat{x}^{B},\hat{p}^{B})^{T}$, which satisfies the canonical commutation relations $[\hat{R}_{k},\hat{R}_{l}]=i\Omega_{kl}$, where $\Omega=
\begin{pmatrix}
0 & 1\\
-1 & 0
\end{pmatrix}^{\oplus 2}$ is the symplectic matrix. The two-mode Gaussian state $\rho_{AB}$ is completely characterized by its first and second statistical moments. The latter define the covariance matrix $\sigma_{ij}=\mathrm{Tr}\left[\{\Delta\hat{R}_{i},\Delta\hat{R}_{j}\}_{+}\rho_{AB}\right]$ \cite{AA1,AA2}, where $\Delta\hat{R}_{i}=\hat{R}_{i}-\langle\hat{R}_{i}\rangle$. The covariance matrix can be written in the block form \cite{AA3}
\begin{align}\label{A22}
\sigma_{AB}=
\begin{pmatrix}
A & C\\
C^{T} & B
\end{pmatrix},
\end{align}
where the $2\times2$ symmetric matrices $A$ and $B$ are the covariance matrices of the reduced single-mode states of Alice and Bob, respectively, while $C$ contains the correlations between the two modes.

A bipartite two-mode Gaussian state with covariance matrix $\sigma_{AB}$ is nonsteerable from Bob to Alice under Gaussian measurements if and only if \cite{L2}
\begin{align}\label{A23}
\sigma_{AB}+i\left(\Omega_{A}\oplus0_{B}\right)\geq0.
\end{align}
Consequently, the state is steerable from Bob to Alice whenever the above condition is violated. Since the steered subsystem $A$ consists of a single mode, the corresponding Gaussian steering measure is given by \cite{L3}
\begin{align}\label{A24}
\mathcal{G}^{B\rightarrow A}(\sigma_{AB})
=
\max\left\{
0,
\frac{1}{2}\ln\frac{\det B}{\det\sigma_{AB}}
\right\}.
\end{align}
Similarly, by interchanging the roles of Alice and Bob, the Gaussian steering from Alice to Bob is given by
\begin{align}\label{A25}
\mathcal{G}^{A\rightarrow B}(\sigma_{AB})
=
\max\left\{
0,
\frac{1}{2}\ln\frac{\det A}{\det\sigma_{AB}}
\right\},
\end{align}
where $A$ and $B$ denote the covariance submatrices introduced in Eq.~(\ref{A22}).

Unlike quantum entanglement, quantum steering is intrinsically directional and generally satisfies $\mathcal{G}^{B\rightarrow A}(\sigma_{AB})\neq\mathcal{G}^{A\rightarrow B}(\sigma_{AB})$. This asymmetry reflects the distinct roles of the steering and steered parties. To quantify the degree of steering asymmetry, we define the steering difference as
\begin{align}\label{A26}
\mathcal{G}^{\Delta}_{AB}
=
\left|
\mathcal{G}^{A\rightarrow B}(\sigma_{AB})
-
\mathcal{G}^{B\rightarrow A}(\sigma_{AB})
\right|.
\end{align}
It has been shown that, for an arbitrary two-mode Gaussian state, the steering asymmetry cannot exceed $\ln 2$.
Whenever Gaussian steering is present, its directional behavior can be classified into two categories. Two-way steering occurs when $\mathcal{G}^{A\rightarrow B}>0$ and $\mathcal{G}^{B\rightarrow A}>0$, whereas one-way steering occurs when $\mathcal{G}^{A\rightarrow B}>0$ and $\mathcal{G}^{B\rightarrow A}=0$, or vice versa. In the following, we derive the conditions under which one-way and two-way Gaussian steering arise in de-Sitter space.

\section{Distribution of Gaussian tripartite steering in de-Sitter space}
\subsection{Reduction of quantum steerability between initially correlated modes}
The three-mode pure Gaussian state shared by Alice, Bob, and Charlie in an asymptotically flat spacetime is characterised by the covariance matrix \cite{AA4,AA5} 
\begin{align}\label{A27}
\sigma_{ABC}(s)=\left(\begin{array}{ccc}
\sigma_{A}(s)&\xi(s)&\xi(s)\\
\xi(s)&\sigma_{B}(s)&\xi(s)\\
\xi(s)&\xi(s)&\sigma_{C}(s)\\
\end{array}\right),
\end{align}
where the submatrices are defined as
\begin{align}\label{A28}
\sigma_{A}(s)=\sigma_{B}(s)=\sigma_{C}(s)=\left(\begin{array}{cc}
m&0\\
0&m\\
\end{array}\right),
\xi(s)=\left(\begin{array}{cc}
g_{1}&0\\
0&g_{2}\\
\end{array}\right).
\end{align}
Here, the parameters $m$, $g_{1}$, and $g_{2}$ depend on the squeezing parameter $s$ as
\begin{align}\label{A29}
m &= \frac{1}{3}\sqrt{4\cosh(4s)+5}, \nonumber\\
g_{1} &= \frac{2\sinh^{2}(2s)+3\sinh(4s)}{3\sqrt{4\cosh(4s)+5}}, \nonumber\\
g_{2} &= \frac{2\sinh^{2}(2s)-3\sinh(4s)}{3\sqrt{4\cosh(4s)+5}}.
\end{align}
For Charlie's mode, we denote the corresponding squeezing amplitude by $\gamma_C\equiv\gamma_p$, where $\gamma_p$ is given in Eq.~(\ref{A21}). After absorbing the phase of $\gamma_C$ into the definitions of the local modes, the Bunch-Davies vacuum for a global observer can be expressed as the two-mode squeezed state
\begin{align}\label{A30}
|0\rangle_{BD}=\sqrt{1-|\gamma_C|^2}\sum_{n=0}^{\infty}|\gamma_C|^n|n\rangle_R|n\rangle_L .
\end{align}
We then introduce the real squeezing parameter $r_C$ through $\tanh r_C=|\gamma_C|$. The Bunch-Davies vacuum can equivalently be generated from the local vacuum as $|0\rangle_{BD}=\hat{U}_{R,L}(r_C)|0\rangle_R|0\rangle_L$, where \[\hat{U}_{R,L}(r_C)=\exp\left[r_C\left(\hat{c}_{R}^{\dagger}\hat{c}_{L}^{\dagger}-\hat{c}_{R}\hat{c}_{L}\right)\right]\] is the two-mode squeezing operator. Using $\tanh r_C=|\gamma_C|$, one has $\cosh r_C=(1-|\gamma_C|^2)^{-1/2}$ and $\sinh r_C=|\gamma_C|(1-|\gamma_C|^2)^{-1/2}$. Hence, in phase space the transformation is represented by the symplectic matrix
\begin{align}\label{A31}
S_{C,\overline{C}}(\gamma_{C})=\frac{1}{\sqrt{1-|\gamma_{C}|^{2}}}\left(\begin{array}{cccc}
1 & 0 & |\gamma_{C}| & 0\\
0 & 1& 0 & -|\gamma_{C}|\\
|\gamma_{C}| & 0 & 1 & 0\\
0 & -|\gamma_{C}| & 0 & 1\\
\end{array}\right),
\end{align}
indicating that the squeezing operation is applied to the bipartite state shared between Charlie and anti-Charlie.

Under this transformation, the mode accessible to Charlie is effectively distributed across two open charts. In other words, an additional mode, denoted by  $\overline{C}$, becomes relevant for an observer confined to the open charts. Consequently, a complete description of the system requires four modes: mode $ A$ described by Alice, mode $B$ described by Bob, mode $C$ observed by Charlie in the $R$ chart, and mode $\overline{C}$ associated with a hypothetical observer, anti-Charlie, restricted to the $L$ chart. The covariance matrix of the entire state is then expressed as 
\begin{align} 
\sigma_{ABC\overline{C}}(s,\gamma_{C})
&= [I_{AB}\oplus S_{C\overline{C}}(\gamma_{C})]
   [\sigma_{ABC}(s)\oplus I_{\overline{C}}]
   [I_{AB}\oplus S_{C\overline{C}}(\gamma_{C})]^{T} \nonumber\\
&= \left(
 \begin{array}{cccc}
\mathcal{\sigma}_{A}(s) & \xi_{AB}(s) & \xi_{AC}(s,\gamma_{C}) &\xi_{A\bar C}(s,\gamma_{C}) \\
\xi^{\sf T}_{AB}(s) & \mathcal{\sigma}_{B}(s) & \xi_{BC}(s,\gamma_{C}) & \xi_{B\bar C}(s,\gamma_{C}) \\
\xi^{\sf T}_{AC}(s,\gamma_{C}) & \xi^{\sf T}_{BC}(s,\gamma_{C}) & \mathcal{\sigma}_{C}(s,\gamma_{C}) & \xi_{C\bar C}(s,\gamma_{C}) \\
\xi^{\sf T}_{A\bar C}(s,\gamma_{C}) & \xi^{\sf T}_{B\bar C}(s,\gamma_{C}) & \xi^{\sf T}_{C\bar C}(s,\gamma_{C}) & \mathcal{\sigma}_{\bar C}(s,\gamma_{C})
\end{array}
\right)
\label{A32}.
\end{align}
The diagonal blocks in Eq.~(\ref{A32}) take the forms
\begin{equation}\label{A33}
\begin{aligned}
\sigma_{C}(s,\gamma_{C}) &= 
\begin{pmatrix}
\frac{m+|\gamma_{C}|^{2}}{1-|\gamma_{C}|^{2}} & 0 \\[2mm]
0 & \frac{m+|\gamma_{C}|^{2}}{1-|\gamma_{C}|^{2}} 
\end{pmatrix},\\[1mm]
\sigma_{\bar{C}}(s,\gamma_{C}) &= 
\begin{pmatrix}
\frac{1+m|\gamma_{C}|^{2}}{1-|\gamma_{C}|^{2}} & 0 \\
0 & \frac{1+m|\gamma_{C}|^{2}}{1-|\gamma_{C}|^{2}} 
\end{pmatrix}.
\end{aligned}
\end{equation}
The off-diagonal correlation blocks take the following forms
\begin{align}\label{A34}
\xi_{AB}(s)&=
\begin{pmatrix}
g_{1} & 0 \\
0 & g_{2}
\end{pmatrix},\nonumber\\
\xi_{AC}(s,\gamma_{C})&=
\xi_{BC}(s,\gamma_{C})=
\begin{pmatrix}
\frac{g_{1}}{\sqrt{1-|\gamma_{C}|^{2}}} & 0 \\
0 & \frac{g_{2}}{\sqrt{1-|\gamma_{C}|^{2}}}
\end{pmatrix},\nonumber\\
\xi_{A\bar{C}}(s,\gamma_{C})&=
\xi_{B\bar{C}}(s,\gamma_{C})=
\begin{pmatrix}
\frac{|\gamma_{C}| g_{1}}{\sqrt{1-|\gamma_{C}|^{2}}} & 0 \\
0 & -\frac{|\gamma_{C}| g_{2}}{\sqrt{1-|\gamma_{C}|^{2}}}
\end{pmatrix},\nonumber\\
\xi_{C\bar{C}}(s,\gamma_{C})&=
\begin{pmatrix}
\frac{|\gamma_{C}| (1+m)}{1-|\gamma_{C}|^{2}} & 0 \\
0 & -\frac{|\gamma_{C}| (1+m)}{1-|\gamma_{C}|^{2}}
\end{pmatrix}.
\end{align}

 Since Charlie in the $R$ chart cannot access the modes located in the causally disconnected $L$ region, it is necessary to trace over these inaccessible modes. By tracing out mode $\overline{C}$ in the $L$ chart, we obtain the covariance matrix $\sigma_{ABC}(s,\gamma_C)$ describing the observable modes shared by Alice, Bob, and Charlie. The resulting covariance matrix for the three accessible modes $A$, $B$, and $C$ is given by

\begin{align}\label{A35}
\sigma_{ABC}(s,\gamma_{C}) &=\!
\left(\begin{array}{ccc}
\sigma_{A}(s) & \xi_{AB}(s) & \xi_{AC}(s,\gamma_{C})\\
\xi_{AB}^{T}(s) & \sigma_{B}(s) & \xi_{BC}(s,\gamma_{C})\\
\xi_{AC}^{T}(s,\gamma_{C}) & \xi_{BC}^{T}(s,\gamma_{C}) & \sigma_{C}(s,\gamma_{C})\\
\end{array}\right).
\end{align}
By employing the covariance matrix $\sigma_{ABC}(s,\gamma_{C})$ together with Eqs.(\ref{A24}) and (\ref{A25}), we obtain the analytical expressions for Gaussian tripartite steering corresponding to different $2\rightarrow1$ and $1\rightarrow2$ bipartitions of the state described by Eq.(\ref{A35}). The resulting steering measures are expressed as:
\begin{align}\label{A36}
\mathcal{G}^{AB \rightarrow\ C}(\sigma_{ABC})= \mathcal{G}^{A \rightarrow BC}(\sigma_{ABC})= \max\left\{0,\frac{1}{2}\ln{\frac{\frac{1}{9}[5+4\cosh(4s)]}{\det\sigma_{ABC}}}\right\},
\end{align}
\begin{align}\label{A37}
\mathcal{G}^{C \rightarrow AB}(\sigma_{ABC}) = \max\left\{ 0, \frac{1}{2} \ln \frac{\det \sigma_C}{\det \sigma_{ABC}} \right\},
\end{align}
\begin{align}\label{A38}
\mathcal{G}^{BC \rightarrow A}(\sigma_{ABC}) = \max\left\{ 0, \frac{1}{2} \ln \frac{\det \sigma_{BC}}{\det \sigma_{ABC}} \right\},
\end{align}
where
\begin{align*}
\det\sigma_{ABC}=\frac{9 + 5 |\gamma_{C}|^4 + 4 |\gamma_{C}|^4 \cosh(4 s) + 6 |\gamma_{C}|^2 \sqrt{5 + 4 \cosh(4 s)}}{9 (|\gamma_{C}|^2 - 1)^2},
\end{align*}
\begin{align*}
\det\sigma_{C}=\frac{5 + 9 |\gamma_C|^4 + 4 \cosh(4 s) + 6 |\gamma_C|^2 \sqrt{5 + 4 \cosh(4 s)}}{9 (|\gamma_C|^2 - 1)^2},
\end{align*}
\begin{align*}
\det \sigma_{BC}
&= \frac{24 |\gamma_C|^{2}+ 5 \sqrt{5 + 4 \cosh(4 s)}+ 5 |\gamma_C|^{4} \sqrt{5 + 4 \cosh(4 s)}}{9 (|\gamma_C|^{2} - 1)^{2} \sqrt{5 + 4 \cosh(4 s)}} \nonumber\\
&\quad
+ \frac{4 \cosh(4 s)\left(7 |\gamma_C|^{2}+ \sqrt{5 + 4 \cosh(4 s)}+ |\gamma_C|^{4} \sqrt{5 + 4 \cosh(4 s)}\right)}{9 (|\gamma_C|^{2} - 1)^{2} \sqrt{5 + 4 \cosh(4 s)}}\nonumber\\
&\quad+ \frac{2 |\gamma_C|^{2} \cosh(8 s)}{9 (|\gamma_C|^{2} - 1)^{2} \sqrt{5 + 4 \cosh(4 s)}}.
\end{align*}
\begin{figure}[htbp]
\centering

\begin{minipage}{0.48\linewidth}
\includegraphics[width=\linewidth,height=0.2\textheight]{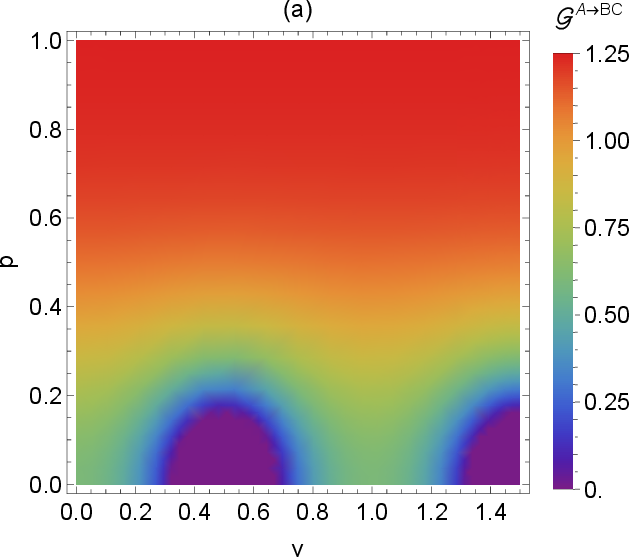}
\end{minipage}\hfill
\begin{minipage}{0.48\linewidth}
\includegraphics[width=\linewidth,height=0.2\textheight]{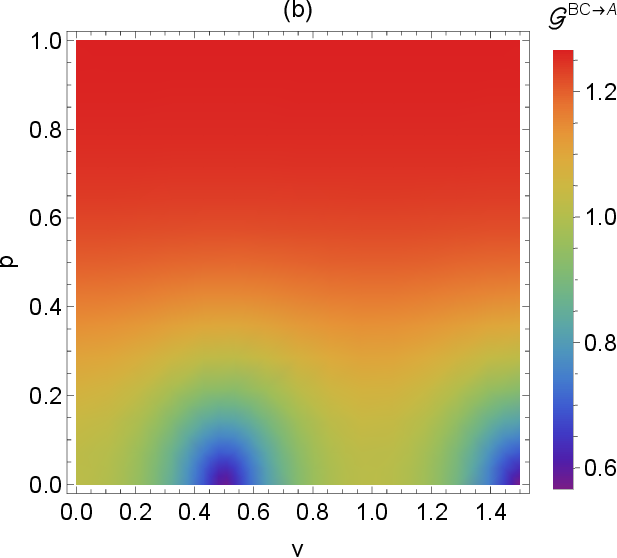}
\end{minipage}

\begin{minipage}{0.48\linewidth}
\includegraphics[width=\linewidth,height=0.2\textheight]{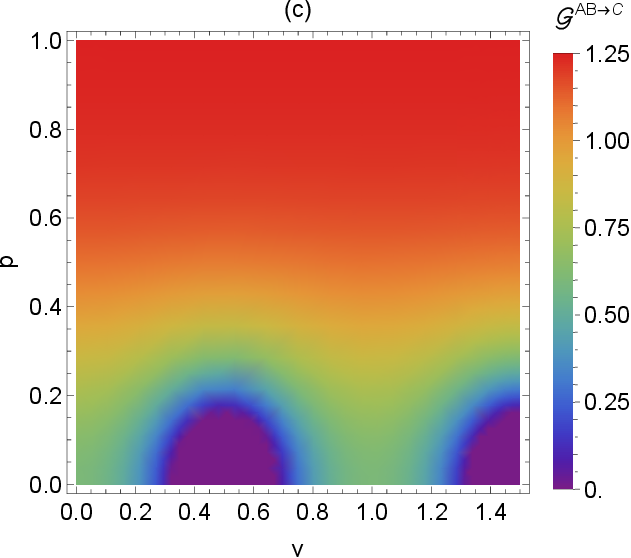}
\end{minipage}\hfill
\begin{minipage}{0.48\linewidth}
\includegraphics[width=\linewidth,height=0.2\textheight]{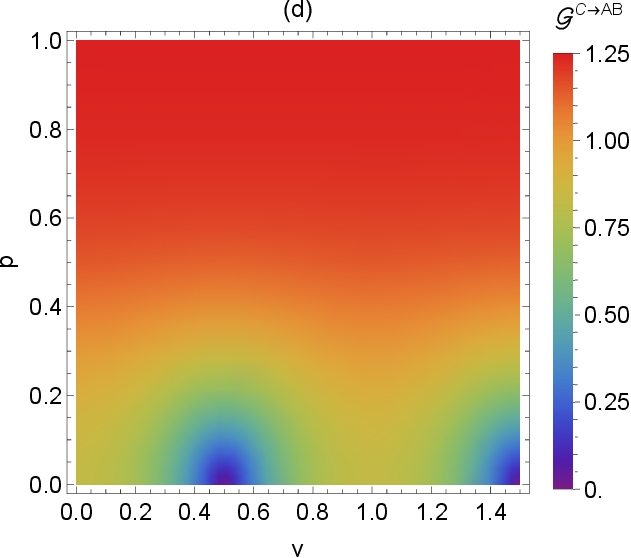}
\end{minipage}

\caption{The Gaussian quantum steerability $\mathcal{G}^{A \rightarrow\ BC}$, $\mathcal{G}^{BC \rightarrow\ A}$, $\mathcal{G}^{AB \rightarrow\ C}$, and $\mathcal{G}^{C \rightarrow\ AB}$ as functions of the curvature parameter $p$ and mass parameter $\nu$. The initial squeezing parameter is fixed at s = 1.0.}
\label{F1}
\end{figure}
From the above expressions, it is evident that Gaussian tripartite steering is determined not only by the initial squeezing parameter $s$, which quantifies the amount of quantum resources encoded in the state, but also by the de-Sitter parameters $p$ and $\nu$, which characterize the space curvature and scalar-field mass, respectively. The dependence on these parameters reflects the sensitivity of Gaussian tripartite steering to gravitational effects in de-Sitter space. It should be noted that, owing to the symmetry between modes $A$ and $B$, only four representative steering configurations need to be considered, namely $\mathcal{G}^{A\rightarrow BC}$, $\mathcal{G}^{BC\rightarrow A}$, $\mathcal{G}^{AB\rightarrow C}$, and $\mathcal{G}^{C\rightarrow AB}$.

As shown in Fig.~{\ref{F1}}, we plot the Gaussian tripartite steering measures in de-Sitter space $\mathcal{G}^{A\rightarrow BC}$, $\mathcal{G}^{BC\rightarrow A}$, $\mathcal{G}^{AB\rightarrow C}$, and $\mathcal{G}^{C\rightarrow AB}$ as functions of the curvature parameter $p$ and the mass parameter $\nu$ for a fixed squeezing parameter $s=1$. It can be seen that all four steering quantities decrease monotonically with decreasing $p$. Since the curvature of de-Sitter space becomes stronger as $p$ approaches zero, this behavior indicates that the de-Sitter gravitational field continuously suppresses the steerable quantum correlations shared among the initial modes. A remarkable feature is that all four steering measures exhibit pronounced local minima around the conformal scalar limit $\nu=\frac{1}{2}$ and the massless scalar limit $\nu=\frac{3}{2}$. This implies that the influence of the de-Sitter background on Gaussian steering is particularly significant in these parameter regimes. In addition, we find that the steering $\mathcal{G}^{A\rightarrow BC}$ undergoes sudden death at a finite curvature when $|\gamma_C|=\sqrt{\frac{\sqrt{5+4\cosh(4s)}-3}
{2\sqrt{5+4\cosh(4s)}}}$. By contrast, the reverse steering $\mathcal{G}^{BC\rightarrow A}$ remains nonzero throughout the entire parameter region. This behavior differs markedly from the bipartite case, where steering associated with observers in the observable region typically undergoes sudden death under sufficiently strong gravitational effects \cite{ACP9}. The persistence of $\mathcal{G}^{BC\rightarrow A}$ and $\mathcal{G}^{AC\rightarrow B}$ demonstrates that tripartite Gaussian steering possesses enhanced robustness against de-Sitter-induced degradation and can preserve quantum resources even in strong gravitational backgrounds. Furthermore, the distinct responses of opposite steering directions reveal the intrinsic asymmetry of Gaussian tripartite steering in de-Sitter space. 
A similar directional dependence is observed for the steering pair 
$\mathcal{G}^{AB\rightarrow C}$ and $\mathcal{G}^{C\rightarrow AB}$. 
While $\mathcal{G}^{AB\rightarrow C}$ undergoes sudden death at a finite 
curvature, the reverse steering $\mathcal{G}^{C\rightarrow AB}$ exhibits 
greater robustness against the de-Sitter gravitational effect. In particular, 
at the conformal scalar limit $\nu=\tfrac{1}{2}$ and the massless scalar limit 
$\nu=\tfrac{3}{2}$, $\mathcal{G}^{C\rightarrow AB}$ remains finite for any 
finite curvature and vanishes only asymptotically in the infinite-curvature 
limit $p\rightarrow0$.

The steering configurations can be classified into three distinct categories 
according to the roles of the gravitational and non-gravitational modes in the 
steering and steered subsystems. First, steering from non-gravitational modes 
to either the gravitational mode or a hybrid gravitational-non-gravitational 
subsystem undergoes sudden death at a finite curvature. Second, when the 
gravitational mode acts as the steering subsystem and the non-gravitational 
modes as the steered subsystem, the steering decreases continuously with 
increasing curvature and eventually vanishes in the strong-curvature regime. 
In sharp contrast, steering from a hybrid gravitational-non-gravitational 
subsystem to a non-gravitational mode remains finite throughout the entire 
parameter region. These three qualitatively distinct behaviors reveal a 
pronounced directional dependence of Gaussian tripartite steering on the roles 
played by the gravitational and non-gravitational modes. The de-Sitter 
gravitational field therefore does not affect all steering partitions 
uniformly, but instead gives rise to a rich and direction-dependent 
redistribution of multipartite steering among different subsystems.

Unlike multipartite entanglement, multipartite  steering is intrinsically direction-dependent and therefore exhibits an inherent asymmetry in de-Sitter space. This distinctive feature has attracted considerable attention in both theoretical and experimental studies \cite{AA6,L10,L12}. As shown in Fig.~{\ref{F1}}, the de-Sitter gravitational effect not only suppresses Gaussian tripartite steering but also induces a noticeable imbalance between opposite steering directions. To quantify this effect, we introduce the Gaussian steering asymmetry as
\begin{align} \label{A39}
\mathcal{G}^{\Delta}_{A|BC}=\left|\mathcal{G}^{A\rightarrow BC}-\mathcal{G}^{BC\rightarrow A}\right|. 
\end{align} 
Similarly, for the steering pair $\mathcal{G}^{AB\rightarrow C}$ and $\mathcal{G}^{C\rightarrow AB}$, we define 
\begin{align} \label{A40}
\mathcal{G}^{\Delta}_{AB|C}=\left|\mathcal{G}^{AB\rightarrow C}-\mathcal{G}^{C\rightarrow AB}\right|. 
\end{align}
These quantities provide a quantitative measure of the intrinsic asymmetry of Gaussian tripartite steering in de-Sitter space and enable us to gain deeper insight into the influence of the de-Sitter gravitational field on multipartite quantum correlations. In order to investigate the behavior of steering asymmetry, it is useful to distinguish between two-way steering and one-way steering. The former corresponds to the situation where steering exists in both directions, whereas the latter occurs when steering is present only in one direction and absent in the reverse direction. Accordingly, the steering asymmetry is discussed separately in the one-way and two-way steering regimes.

\begin{figure}[!t]
\centering
\includegraphics[width=0.48\textwidth]{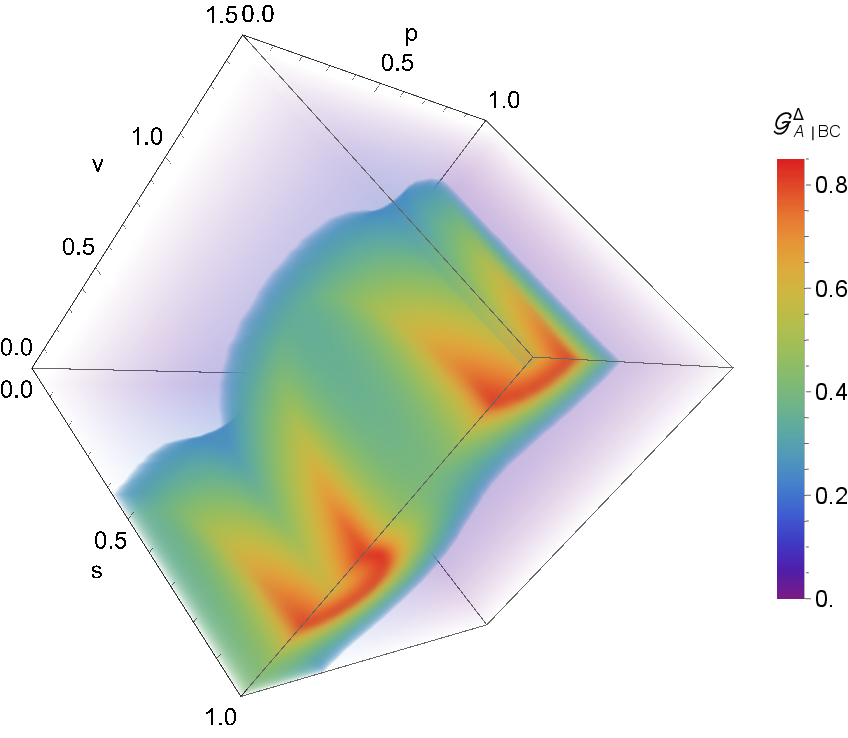}
\hfill
\includegraphics[width=0.48\textwidth]{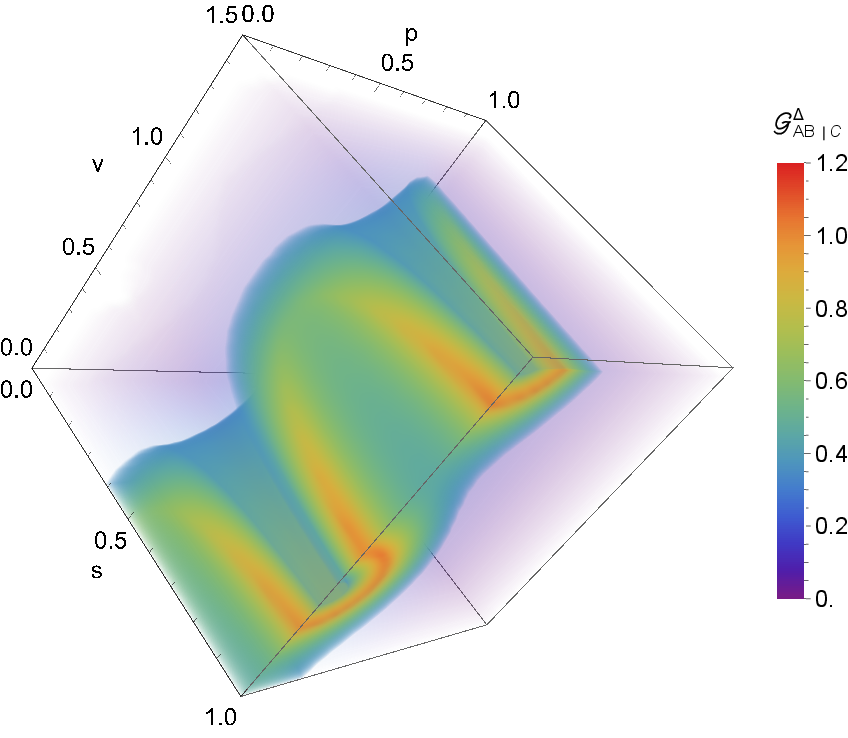}
\caption{The Gaussian tripartite steering asymmetry $\mathcal{G}^{\Delta}_{A|BC}$ and $\mathcal{G}^{\Delta}_{AB|C}$  as functions
of the curvature parameter $p$, mass parameter $\nu$, and the initial squeezing parameter  $s$.}\label{F2}
\end{figure}

Fig.~{\ref{F2}} shows the Gaussian steering asymmetries $\mathcal{G}^{\Delta}_{A|BC}$ and $\mathcal{G}^{\Delta}_{AB|C}$ as functions of the curvature parameter $p$, the mass parameter $\nu$, and the squeezing parameter $s$ in de-Sitter space. Since the system can exhibit both one-way and two-way steering in different parameter regions, the corresponding steering asymmetries are described by different analytical expressions. In the one-way steering regime, where only $\mathcal{G}^{BC\rightarrow A}$ remains nonvanishing, the asymmetry is given by
\begin{align}\label{A41}
\mathcal{G}^{\Delta1}_{A|BC}=\max\left\{0,\,\frac{1}{2}\ln\frac{\det\sigma_{BC}}{\det\sigma_{ABC}}\right\}.
\end{align}
In the two-way steering regime, where both steering directions coexist, the asymmetry becomes
\begin{align}\label{A42}
\mathcal{G}^{\Delta2}_{A|BC}=\max\left\{0,\,\frac{1}{2}\ln\frac{\det\sigma_{BC}}{\frac{1}{9}\left[5+4\cosh(4s)\right]}\right\}.
\end{align}
Similarly, in the one-way steering region, the Gaussian steering asymmetry 
$\mathcal{G}^{\Delta}_{AB|C}$ is expressed as
\begin{align}\label{A43}
\mathcal{G}^{\Delta1}_{AB|C}=\max\left\{0,\,\frac{1}{2}\ln\frac{\det\sigma_{C}}{\det\sigma_{ABC}}\right\},
\end{align}
whereas in the two-way steering region it reduces to
\begin{align}\label{A44}
\mathcal{G}^{\Delta2}_{AB|C}=\max\left\{0,\,\frac{1}{2}\ln\frac{\det\sigma_{C}}{\frac{1}{9}\left[5+4\cosh(4s)\right]}\right\}.
\end{align}
In Fig.~{\ref{F2}}, we observe that, for a fixed $\nu$, the curvature parameter $p$ corresponding to the maximal steering asymmetry decreases as the squeezing parameter $s$ increases. We further observe that the asymmetries reach their maxima in the parameter regions where $\mathcal{G}^{A\rightarrow BC}$ and $\mathcal{G}^{AB\rightarrow C}$ undergo sudden death, corresponding to the transition from two-way steering to one-way steering. These results indicate that the de-Sitter gravitational field not only suppresses Gaussian tripartite steering but also significantly enhances its intrinsic asymmetry.

\subsection{Gaussian tripartite steering involving the inaccessible mode}
\begin{figure}[!t]
\centering

\begin{minipage}[t]{0.32\textwidth}
    \centering
    \includegraphics[width=\linewidth,height=4.0cm]{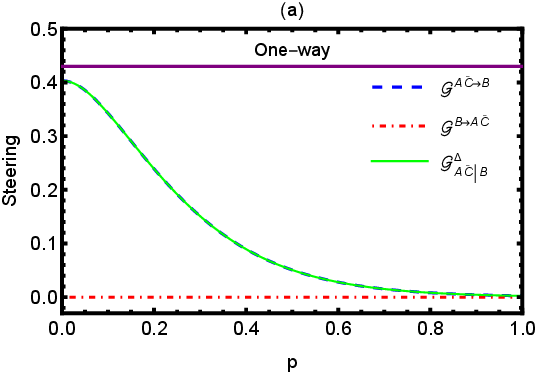}
\end{minipage}
\hfill
\begin{minipage}[t]{0.32\textwidth}
    \centering
    \includegraphics[width=\linewidth,height=4.0cm]{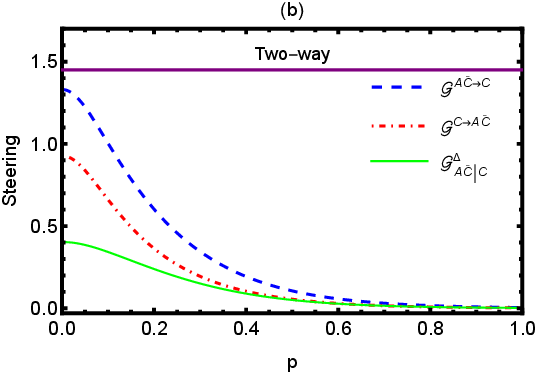}
\end{minipage}
\hfill
\begin{minipage}[t]{0.32\textwidth}
    \centering
    \includegraphics[width=\linewidth,height=4.0cm]{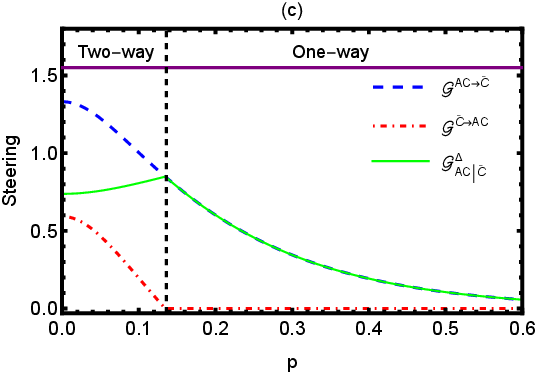}
\end{minipage}

\caption{Gaussian tripartite steering and steering asymmetries  between initially uncorrelated modes as functions of the curvature parameter $p$. The initial squeezing parameter is fixed at $s=1$, and the mass parameter is set to $\nu=0.4$.}
\label{F3}
\end{figure}

In this subsection, we investigate Gaussian tripartite steering involving the inaccessible mode $\bar{C}$ in de-Sitter space. The corresponding covariance matrices and analytical expressions for all Gaussian steering measures are given in Appendix A. We now focus on the directional properties of the Gaussian tripartite steering involving the inaccessible region in de-Sitter space, as shown in Fig.~{\ref{F3}}.

In Fig.~{\ref{F3}}, we plot the Gaussian tripartite steering involving the causally disconnected mode $\bar{C}$, namely $\mathcal{G}^{A\bar{C}\rightarrow B}$, $\mathcal{G}^{B\rightarrow A\bar{C}}$, $\mathcal{G}^{A\bar{C}\rightarrow C}$, $\mathcal{G}^{C\rightarrow A\bar{C}}$, $\mathcal{G}^{AC\rightarrow\bar{C}}$, and $\mathcal{G}^{\bar{C}\rightarrow AC}$, as functions of the de-Sitter curvature parameter $p$. From Fig.4, it can be seen that the de-Sitter gravitational field gives rise to three distinct directional behaviors of Gaussian tripartite steering involving the causally disconnected mode. (i) For subsystem $B$ and the collective subsystem $A\bar C$, the de-Sitter space curvature cannot generate steering in the direction $B\rightarrow A\bar C$, whereas steering can be generated in the reverse direction $A\bar C\rightarrow B$. Moreover, the steering in the latter direction is continuously enhanced as the curvature parameter $p$ decreases, giving rise to persistent one-way steering. (ii) For subsystem $C$ and the collective subsystem $A\bar C$, the de-Sitter space curvature can generate steering in both directions. Furthermore, the steering in the direction $A\bar C\rightarrow C$ is always stronger than that in the reverse direction $C\rightarrow A\bar C$, resulting in persistent asymmetric two-way steering. (iii) For subsystem $\bar C$ and the collective subsystem $AC$, the de-Sitter space curvature can generate steering in the direction $AC\rightarrow\bar C$, whereas steering in the reverse direction $\bar C\rightarrow AC$ cannot be generated in the weak-curvature regime. As the curvature parameter $p$ decreases, the steering in the direction $\bar C\rightarrow AC$ undergoes sudden birth at a critical curvature parameter. This sudden-birth point marks the transition from one-way steering to two-way steering and simultaneously corresponds to the maximum steering asymmetry. These results demonstrate that the de-Sitter space curvature not only generates Gaussian tripartite steering involving the causally inaccessible mode $\bar C$, but also leads to diverse steering characteristics depending on the steering direction. In particular, the emergence of persistent one-way steering, persistent asymmetric two-way steering, and the transition from one-way to two-way steering reveals the rich directional structures of Gaussian tripartite steering in de-Sitter space, highlighting the crucial role of the de-Sitter gravitational effect in redistributing steering resources between observable and causally disconnected regions.

\section{Conclusions \label{GSCDGE}}

In this work, we have investigated the sharing, redistribution, and directional
asymmetry of Gaussian tripartite quantum steering in de-Sitter space within
the continuous-variable framework. By expressing the Bunch-Davies vacuum in
terms of the open-chart modes through a Bogoliubov transformation, we have
derived the covariance matrices of the resulting multipartite Gaussian states
and analyzed both one-to-two and two-to-one steering configurations. For the
initially correlated modes in the physically accessible region, the de-Sitter
gravitational field generally suppresses Gaussian tripartite steering, while
the degradation exhibits a pronounced dependence on the steering direction.
Depending on the roles played by the gravitational and non-gravitational modes
in the steering and steered subsystems, three qualitatively distinct behaviors
emerge: sudden death at a finite curvature, continuous suppression toward
vanishing steering in the strong-curvature regime, and persistent survival
throughout the entire parameter region. In particular, the persistence of
two-to-one steering toward a non-gravitational mode demonstrates that
tripartite steering can exhibit stronger robustness against de-Sitter-induced
degradation than its bipartite counterpart  \cite{ACP9,ACP10,ACP11,ACP12}. We have also found that the
steering asymmetry reaches its maximum at the critical point where one
steering direction undergoes sudden death, identifying the transition from
two-way to one-way steering in de-Sitter space. These results show that the de-Sitter
gravitational field does not affect different multipartite steering partitions
uniformly, but instead gives rise to a strongly direction-dependent
redistribution of steerable quantum correlations.

More remarkably, the de-Sitter gravitational field can generate Gaussian
tripartite steering involving the causally disconnected mode $\bar{C}$, even
when the corresponding modes are initially uncorrelated. The generated
steering displays rich directional structures depending on the partition of
the steering and steered subsystems. We have identified persistent one-way
steering, persistent asymmetric two-way steering, and a sudden-birth
transition from one-way to two-way steering. In the latter case, the onset of
the newly generated steering coincides with the maximum steering asymmetry,
revealing a direct connection between the emergence of a new steering
direction and the directional imbalance of multipartite quantum correlations.
Taken together, our results demonstrate that de-Sitter curvature plays a dual
role in the dynamics of multipartite quantum steering: it not only degrades
preexisting steering in the observable sector, but also redistributes and
generates directional steering across the cosmological horizon. The pronounced
directional dependence and enhanced robustness revealed here highlight the
richer structure of multipartite steering beyond bipartite correlations and
provide further insight into the behavior of quantum resources in de-Sitter space.

\begin{acknowledgments}
This work is supported by the National Natural Science Foundation of China (12575056),   the Natural
Science Foundation of the Liaoning Scientific Committee No. 2026-MS-281, and LiaoNing Revitalization Talents Program (XLYC2503099).	
\end{acknowledgments}

\appendix

\section{Gaussian tripartite steering involving the inaccessible region}
\label{app:steering}
In this appendix, we provide the covariance matrices and analytical expressions 
for Gaussian tripartite steering involving the inaccessible mode $\bar C$ in 
region $L$. Since the modes $C$ and $\bar C$ belong to the causally disconnected 
regions $R$ and $L$, respectively, different reduced tripartite states can be 
obtained by tracing over selected modes. This allows us to characterize the 
redistribution of Gaussian steering between the physically accessible and 
inaccessible sectors.

We first trace over mode $C$ in region $R$. The resulting reduced state is defined on the tripartite subsystem $(A,B,\bar C)$, whose covariance matrix is given by
\begin{align}
\sigma_{AB\bar{C}}(s,\gamma_{C})=\left(
\begin{array}{ccc}
\sigma_{A}(s) & \xi_{AB}(s) & \xi_{A\bar{C}}(s,\gamma_{C})\nonumber\\
\xi_{AB}^{T}(s) & \sigma_{B}(s) & \xi_{B\bar{C}}(s,\gamma_{C})\nonumber\\
\xi_{A\bar{C}}^{T}(s,\gamma_{C}) & \xi_{B\bar{C}}^{T}(s,\gamma_{C}) & \sigma_{\bar{C}}(s,\gamma_{C})
\end{array}\right).
\end{align}
Here, the diagonal blocks describe the individual modes, while the off-diagonal 
blocks characterize the correlations between them. For this reduced state, 
three steering configurations vanish identically,
\begin{align}\label{AA1}
\mathcal{G}^{AB\rightarrow \bar{C}}=\mathcal{G}^{\bar{C}\rightarrow AB}=\mathcal{G}^{B\rightarrow A\bar{C}}=\mathcal{G}^{A\rightarrow B\bar{C}}=0.
\end{align}
Therefore, the joint subsystem $AB$ cannot steer the inaccessible mode $\bar C$, nor can $\bar C$ steer the joint subsystem $AB$. Moreover, mode $B$ alone cannot steer the joint subsystem $A\bar C$. 
The steering from the joint subsystem $B\bar C$ to mode $A$ is equal to that from $A\bar C$ to mode $B$ and is given by
\begin{align}\label{AA3}
\mathcal{G}^{B\bar{C}\rightarrow A}=\mathcal{G}^{A\bar{C}\rightarrow B}=
\max\left\{
0,
\frac{1}{2}\ln
\frac{\det \sigma_{B\bar C}}
{\det\sigma_{AB\bar C}}
\right\}.
\end{align}
The determinants entering the above steering measures are
\begin{align*}
\det \sigma_{A\bar C}
=\det \sigma_{B\bar C}
&=\frac{24 |\gamma_C|^2+5\sqrt{5+4\cosh(4s)}
+5|\gamma_C|^4\sqrt{5+4\cosh(4s)}}
{9(|\gamma_C|^2-1)^2\sqrt{5+4\cosh(4s)}}\nonumber\\
&\quad+\frac{4\cosh(4s)
\left[7|\gamma_C|^2+\sqrt{5+4\cosh(4s)}
+|\gamma_C|^4\sqrt{5+4\cosh(4s)}\right]}
{9(|\gamma_C|^2-1)^2\sqrt{5+4\cosh(4s)}}\nonumber\\
&\quad+\frac{2|\gamma_C|^2\cosh(8s)}
{9(|\gamma_C|^2-1)^2\sqrt{5+4\cosh(4s)}},
\end{align*}
and
\begin{align*}
\det\sigma_{AB\bar C}=
\frac{5+9|\gamma_C|^4+4\cosh(4s)
+6|\gamma_C|^2\sqrt{5+4\cosh(4s)}}
{9(|\gamma_C|^2-1)^2}.
\end{align*}

We next trace over mode $B$ and retain the tripartite subsystem 
$(A,C,\bar C)$. This reduced state allows us to characterize the redistribution 
of Gaussian steering involving the observable mode $C$ in region $R$ and its 
inaccessible partner $\bar C$ in region $L$. The corresponding covariance 
matrix is
\begin{align}
\sigma_{AC\bar C}(s,\gamma_C)=\left(
\begin{array}{ccc}
\sigma_A(s)&\xi_{AC}(s,\gamma_C)&\xi_{A\bar C}(s,\gamma_C)\nonumber\\
\xi_{AC}^{T}(s,\gamma_C)&\sigma_C(s,\gamma_C)&\xi_{C\bar C}(s,\gamma_C)\nonumber\\
\xi_{A\bar C}^{T}(s,\gamma_C)&\xi_{C\bar C}^{T}(s,\gamma_C)&\sigma_{\bar C}(s,\gamma_C)
\end{array}\right).
\end{align}
Here, the diagonal blocks describe the individual modes, while the off-diagonal 
blocks characterize the correlations between them. For this reduced state, 
Gaussian steering between mode $A$ and the joint subsystem $C\bar C$ vanishes 
in both directions,
\begin{align}\label{AA4}
\mathcal{G}^{A\rightarrow C\bar C}=
\mathcal{G}^{C\bar C\rightarrow A}
=0.
\end{align}
By contrast, nonvanishing steering arises for other bipartitions involving the 
observable and inaccessible modes. In particular, the steering measures are given by
\begin{align}\label{AA5}
\mathcal{G}^{A\bar C\rightarrow C}= \mathcal{G}^{AC\rightarrow \bar C}=\max \left\{
0,
\frac12\ln
\frac{\det\sigma_{AC}}
{\frac19(5+4\cosh(4s))}
\right\}.
\end{align}
The steering from the inaccessible mode $\bar C$ to the joint subsystem $AC$ 
takes the form
\begin{align}\label{AA6}
\mathcal{G}^{\bar C\rightarrow AC}=
\max\left\{
0,
\frac12\ln
\frac{9+5|\gamma_C|^4+4|\gamma_C|^4\cosh(4s)
+6|\gamma_C|^2\sqrt{5+4\cosh(4s)}}
{(1-|\gamma_C|^2)^2(5+4\cosh(4s))}
\right\}.
\end{align}
For the partition \(C|A\bar C\), the steering from the observable mode \(C\) to the joint subsystem \(A\bar C\) is given by
\begin{align}\label{AA7}
\mathcal{G}^{C\rightarrow A\bar C}=
\max\left\{
0,
\frac12\ln
\frac{5+9|\gamma_C|^4+4\cosh(4s)
+6|\gamma_C|^2\sqrt{5+4\cosh(4s)}}
{(1-|\gamma_C|^2)^2(5+4\cosh(4s))}
\right\}.
\end{align}
The distinct analytical forms of 
$\mathcal{G}^{\bar C\rightarrow AC}$ and 
$\mathcal{G}^{C\rightarrow A\bar C}$ illustrate the dependence of Gaussian steering on the choice of the steering and steered subsystems. The determinants entering the above expressions satisfy
\begin{align}\label{AA8}
\det\sigma_{AC}=\det\sigma_{BC}=
\det\sigma_{A\bar C}.
\end{align}
These analytical results characterize the steering properties of the reduced 
state $\sigma_{AC\bar C}$ and provide the basis for analyzing the redistribution 
and directional asymmetry of Gaussian tripartite steering across the 
cosmological horizon.

\end{document}